# Canalization-based super-resolution imaging using a single van der Waals layer

Jiahua Duan[1,2,3,4,&,*], Aitana Tarazaga Martin-Luengo[3,&,*], Christian Lanza[3,&], Stefan Partel[5], Kirill Voronin[6], Ana Isabel F. Tresguerres-Mata[3], Gonzalo Álvarez-Pérez[3,4], Alexey Y. Nikitin[6,7], J. Martín-Sánchez[3,4], P. Alonso-González[3,4,*]

[1] Centre for Quantum Physics, Key Laboratory of Advanced Optoelectronic Quantum Architecture and Measurement (MOE), School of Physics, Beijing Institute of Technology, Beijing, China.
[2] Beijing Key Laboratory of Nanophotonics and Ultrafine Optoelectronic Systems, Beijing Institute of Technology, Beijing, China.
[3] Department of Physics, University of Oviedo, Oviedo, Spain.
[4] Center of Research on Nanomaterials and Nanotechnology, CINN (CSIC-Universidad de Oviedo), El Entrego, Spain.
[5] Vorarlberg University of Applied Sciences, Research Center of Microtechnology, Austria.
[6] Donostia International Physics Center (DIPC), Donostia, San Sebastian, Spain.
[7] IKERBASQUE, Basque Foundation for Science, Bilbao, Spain.
[&]These authors contributed equally: J. Duan, A. Tarazaga Martin-Luengo, C. Lanza
Emails: duanjiahua@bit.edu.cn, aitanatarazaga@uniovi.es, pabloalonso@uniovi.es

## Abstract

Canalization is an optical phenomenon that enables unidirectional propagation of light in a natural way, i.e., without the need of predefined waveguiding designs. Predicted years ago, it has recently been demonstrated using highly confined phonon polaritons (PhPs) in twisted layers of the van der Waals (vdW) crystal α-MoO$_3$, offering unprecedented possibilities for controlling light-matter interactions at the nanoscale. However, despite this finding, applications based on polariton canalization have remained elusive so far, which can be explained by the complex sample fabrication of twisted stacks. In this work, we introduce a novel canalization phenomenon, arising in a single vdW thin layer (α-MoO$_3$) when it is interfaced with a substrate exhibiting a given negative permittivity, that allows us to demonstrate a proof-of-concept application based on polariton canalization: super-resolution (up to ~λ$_0$/220) nanoimaging. Importantly, we find that canalization-based imaging transcends conventional projection constraints allowing the super-resolution images to be obtained at any desired location in the image plane. This versatility stems from the synergetic manipulation of three distinct parameters: incident frequency, rotation angle of the thin vdW layer, and its thickness. These results not only provide valuable insights into the fundamental properties of canalization, but also constitute a seminal step towards multifaceted photonic applications, encompassing imaging, data transmission and ultra-compact photonic integration.

Phonon polaritons (PhPs)[1-5] — hybrid excitations of photons and crystal lattice vibrations— hold promises for unprecedented control of the flow of energy at the nanoscale[6-9] due to their deep subwavelength confinement combined with ultra-low losses. Furthermore, the recent discovery of PhPs in strongly anisotropic van der Waals (vdW) materials[10-14], has allowed complementing these extraordinary properties with highly directional propagation. A prominent example is PhPs with in-plane hyperbolic propagation in molybdenum trioxide ($\alpha$-MoO$_3$), where the polaritonic isofrequency contour (IFC) — a slice of the dispersion in momentum-frequency space defined by a plane of constant frequency— results in a hyperbolic-like curve[4,10,11]. More interestingly, the hybridization of these PhPs in twisted $\alpha$-MoO$_3$ bilayers[15-18] and trilayers[19] results in an exotic type of diffraction-less propagation, termed polariton canalization, with the potential for on-demand control of nanolight. Yet, the implementation of PhP canalization for applications remains elusive, as twisted vdW stacks necessitate intricate fabrication and precise in-situ control of twist angles, presenting significant technical challenges. Extremely anisotropic propagation has also been predicted[20] and observed[21] in metasurfaces, particularly at frequencies close to the optical phonon (OP). However, metasurface fabrication presents its own set of sophisticated technical challenges, adding a layer of complexity often concomitant with a poor crystalline quality that curtails effective propagation.

In this work, we propose theoretically and demonstrate experimentally a novel canalization phenomenon for PhPs that takes place in single unstructured $\alpha$-MoO$_3$ layers when they are placed on a substrate with a given negative permittivity (SiC). This single-layer polariton canalization exhibits extraordinary properties, such as low losses (PhPs lifetimes of ~2 picoseconds) and a surprisingly broad spectral range (~50 cm$^{-1}$), allowing us to demonstrate nano-imaging of buried nanostructures (Au disks) with deep subdiffractional resolution up to ~$\lambda_0$/220, where $\lambda_0$ designates the free-space wavelength. Furthermore, we find that this nano-imaging scheme, predicated on manipulating parameters like the incident frequency, the rotation angle of the thin layer and its thickness, possesses the capability to project the super-resolution images at specific predetermined locations on the top surface, holding promises for nanophotonic applications in e.g., information transfer, data communication and energy transmission.

We first introduce theoretically the phenomenon of low-loss and broad-band single-layer canalization of PhPs. To do this, we examine how substrates impact the PhPs propagation within individual $\alpha$-MoO$_3$ layers. In particular, we perform analytical calculations combined with comprehensive numerical simulations in which we model a thin (thickness d=150 nm) layer of $\alpha$-MoO$_3$ on different substrates (illustrated in Fig.1a) and consider a vertically oriented electric dipole acting as a polaritonic source. A central tenet of our study is the quantitative evaluation of a potential canalization, which can be described by the flatness of the PhPs IFC[15-19] (denoted by F). Analytically, we determine F by calculating the derivative of the PhP wavevector with respect to the in-plane angle: $F = \frac{dq_x}{d\varphi}$ (see details in Supplementary Information), where $q_x$ represents the x-component of the PhP wavevector, and $\varphi$ is the angle formed between the PhP wavevector and the x-axis (selected along the [100] crystal direction of $\alpha$-MoO$_3$, as shown in the inset of Fig.1b). In scenarios of ideal PhP canalization in individual $\alpha$-MoO$_3$ layers, the IFC would display twin vertical lines aligned with the [001] crystal direction (y-axis), corresponding to a consistent zero value of F, implying an invariant $q_x$ irrespective of shifts in $\varphi$. Given that the low-momentum components dominate the PhP propagation due to the pronounced losses in their high-momentum counterparts, we calculate F for the in-plane angle $\varphi_F = 15°$. Subsequently, this furnishes an analytical framework to discern the optimal substrate permittivity ($\varepsilon_{optimal}$) requisite for PhP canalization in α-MoO₃ layers (see details in Supplementary Information):

$$\varepsilon_{optimal} = -\sqrt{\varepsilon_z[(\varepsilon_y - \varepsilon_x)\cos^2\varphi_F + \varepsilon_y]}$$

where $\varepsilon_x$, $\varepsilon_y$, $\varepsilon_z$ represent the permittivity components along the [100], [001] and [010] crystal direction of α-MoO₃, respectively. Fig.1b (red line) shows $\varepsilon_{optimal}$ as a function of the incident frequency ω covering the hyperbolic reststrahlen band (RB) of α-MoO₃. For comparison, the permittivity of commonly employed substrates[10,22,23] in nano-optics, such as SiO₂ (black line) and Au (orange line), are also plotted in the same spectral range. A clear deviation between the curves is observed, resulting in hyperbolic IFCs (black and orange insets in Fig.1b) for PhPs in α-MoO₃ layers placed on these substrates[24,25]. However, the permittivity of SiC ($\varepsilon_{SiC}$, green line in Fig.1b) aligns closely with $\varepsilon_{optimal}$ within this spectral range. As a result, the IFC at ω = 885 cm⁻¹ of PhPs in a layer of α-MoO₃ placed on SiC (α-MoO₃/SiC) is characterized by two approximately vertical lines (green in Fig.1b) rather than a hyperbola. Such flattening of the IFC in α-MoO₃/SiC suggests the emergence of PhPs canalization. To better understand this result, we juxtapose the analytical dispersion of PhPs in α-MoO₃/SiC (Fig.1c) with that of PhPs in a layer of α-MoO₃ placed on SiO₂ (α-MoO₃/SiO₂, Fig.1d). As shown in Fig.1d, α-MoO₃/SiO₂ supports both the lowest-order (L=0) and high-order (L=1) polaritonic modes along the [100] crystal direction of α-MoO₃, while there are no polaritonic modes along the [001] direction, constituting therefore a prototypical hyperbolic dispersion[10]. This is more clearly visualized in the simulated images of the real part of the electric field along the z direction, $Re(E_z(x, y))$, at the representative frequencies $\omega_1$ = 915 cm⁻¹ (Fig.1i), $\omega_2$ = 885 cm⁻¹ (Fig.1j), $\omega_3$ = 865 cm⁻¹ (Fig.1k), showing in all cases PhPs propagating within hyperbolic sectors centered along the [100] crystal direction (Fig.1i-1k). Below the longitudinal optical (LO) phonon frequency, $\omega_{LO}$ = 850 cm⁻¹, (gray region in Fig.1d) $\varepsilon_x$ and $\varepsilon_y$ are both negative and thus polaritonic modes exist in both the [100] and [001] directions indicating an elliptic dispersion. At the topological transition from an open (hyperbolic) to a closed (elliptic) IFC, i.e., at $\omega_{LO}$, the PhPs electric field is mostly concentrated within a narrow section centered along the [100] direction (Fig.1l). Note that this behavior of anisotropic propagation stems from the α-MoO₃ permittivity taking values of $\varepsilon_y \approx 0$ and $\varepsilon_x < 0$, and thus is not related directly to the substrate employed.

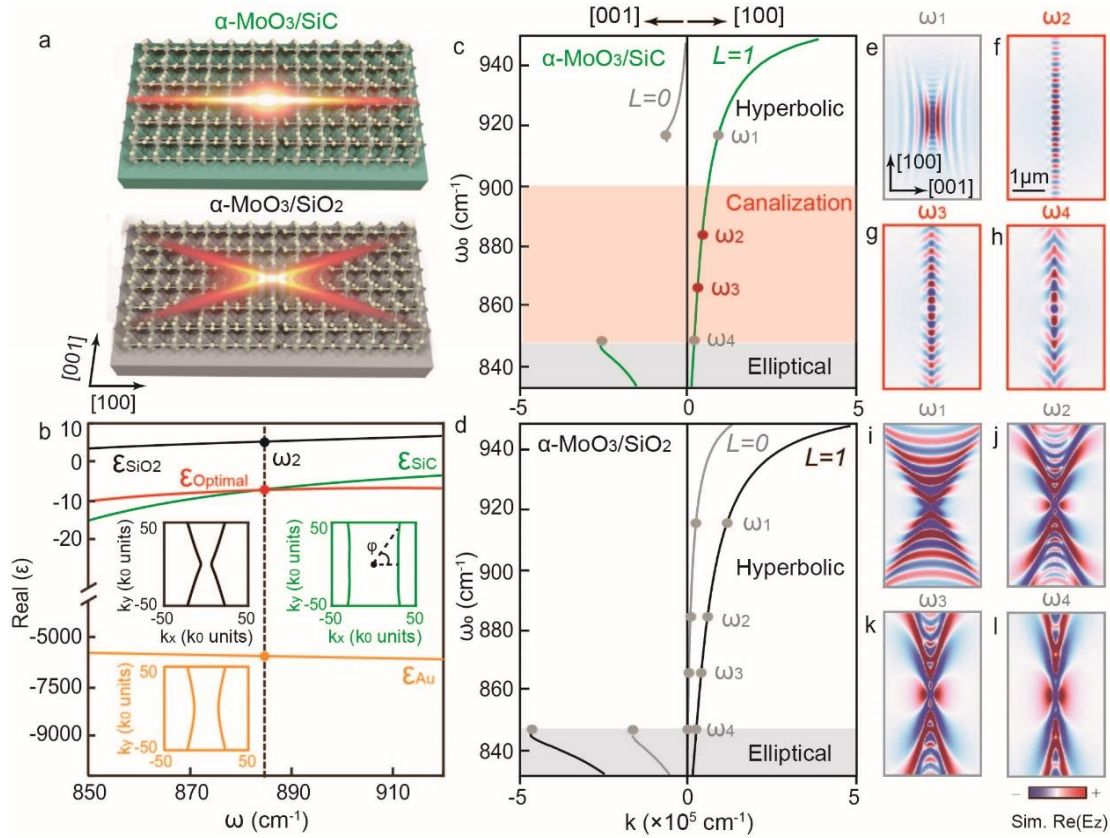

**Figure 1 Low-loss and broadband phonon polariton canalization in individual α-MoO$_3$ layers**. **a,** Schematic of hyperbolic and diffraction-less propagation of phonon polaritons (PhPs) in α-MoO$_3$/SiO$_2$ and α-MoO$_3$/SiC, respectively. **b,** The real component of the permittivity for SiO$_2$ (black solid line), SiC (green solid line), and Au (orange solid line). The red line represents the analytically calculated substrate permittivity ($\varepsilon_{optimal}$) providing an optimal PhP canalization in individual α-MoO$_3$ layers. Inset: analytical PhP IFC for α-MoO$_3$ on the substrate of SiO$_2$ (black square), Au (orange square) and SiC (green square), evaluated at an incident frequency of $\omega_2$ = 885 cm$^{-1}$. **c-d,** Analytical dispersion for PhPs in a 150-nm-thick α-MoO$_3$ layer on SiC (c) and SiO$_2$ (d) along both the [100] and [001] crystal directions. The white, red, and grey shaded regions indicate a hyperbolic, canalized, and elliptical dispersion, respectively. The lines denoted by L=0 and L=1 represent the dispersion of the fundamental (L=0) and high-order (L=1) polaritonic modes in α-MoO$_3$ layers. **e-h,** Simulated near-field distribution, Re[E$_z$(x, y)], excited by a point dipole on a 150-nm-thick α-MoO$_3$ flake on top of SiC at $\omega_1$ = 915 cm$^{-1}$ (e), $\omega_2$ = 885 cm$^{-1}$ (f), $\omega_3$ = 865 cm$^{-1}$ (g) and $\omega_4$ = 850 cm$^{-1}$ (h). **i-l,** Same as (e-h) but for a 150-nm-thick α-MoO$_3$ flake on top of SiO$_2$.

In stark contrast, when a thin α-MoO$_3$ layer is placed on a SiC substrate, the polaritonic dispersion undergoes a dramatic change (Fig.1c) characterized by two distinct effects: i) the emergence of the high-order (L=1) polaritonic mode as the lowest-energy (fundamental) mode along the [100] crystal direction within the hyperbolic RB, which makes the IFC flatter in contrast to an L=0 mode, and ii) the emergence of the fundamental hyperbolic mode (L=0) only along the [001] crystal direction. This result is better analyzed by performing numerical simulations at different frequencies along the dispersion curve. At $\omega_1$ = 915 cm$^{-1}$, the resulting image (Fig.1e) shows hyperbolic propagation of PhPs along the [001] crystal direction of α-MoO$_3$, i.e., the previously forbidden direction for PhPs propagation in thin layers of α-MoO$_3$ on a SiO$_2$ substrate.

This observation has been reported previously[26,27] and explained by the fact that Re($\varepsilon_{SiC}$) reaches a value of -1 at the surface optical (SO) phonon frequency $\omega_{SO} = 943$ cm$^{-1}$. Interestingly, a very different propagation is observed at frequencies $\omega_2 = 885$ cm$^{-1}$ (Fig.1f), $\omega_3 = 865$ cm$^{-1}$ (Fig.1g) and $\omega_4 = 850$ cm$^{-1}$ (Fig.1h), at which PhPs appear confined (collimated) along the [100] crystal direction. This observation demonstrates thus the emergence of a canalization regime of PhPs in a single layer of $\alpha$-MoO$_3$, which, moreover, is shown to occur not only at a single frequency but at several different frequencies, covering the spectral band where $\varepsilon_{SiC}$ almost coincides with $\varepsilon_{optimal}$ (red shaded region in Fig.1c).

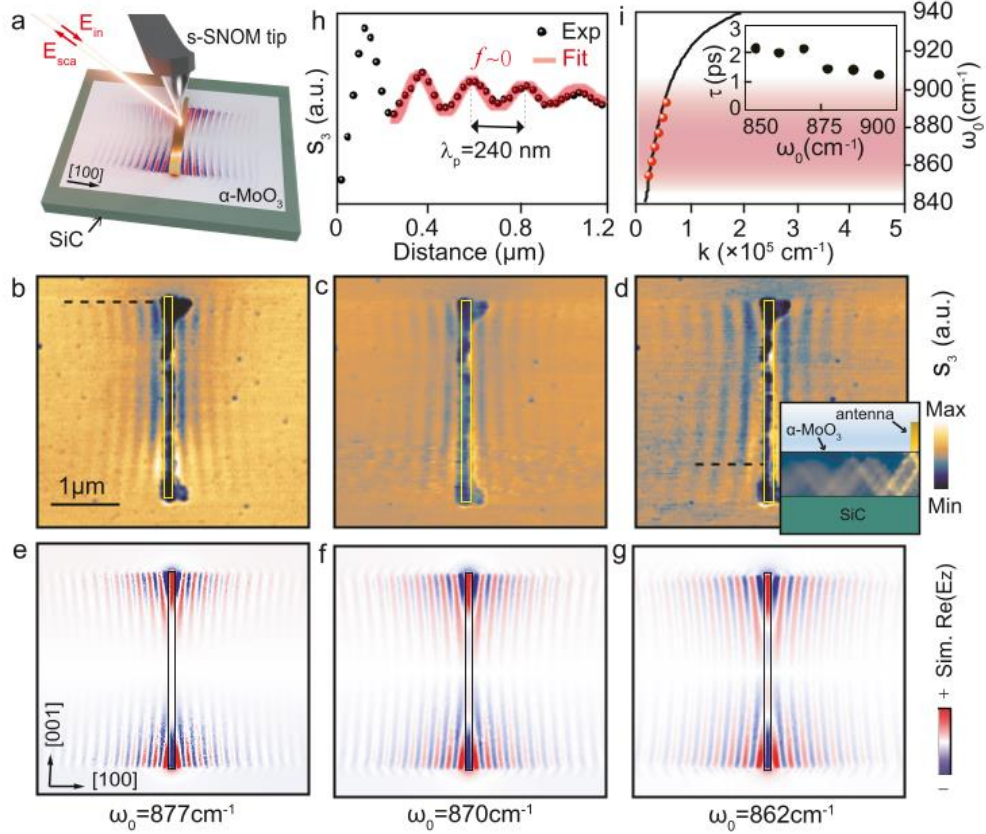

**Figure 2 Visualization of low-loss and broadband PhP canalization in individual $\alpha$-MoO$_3$ layers. a,** Illustration of antenna-launched PhPs in an $\alpha$-MoO$_3$/SiC heterostructure. The spatial distribution of the near field (shown in red and blue) is adapted from the simulation of the near-field distribution, Re[E$_z$ (x, y)]. The s-SNOM tip and sample are simultaneously illuminated with s-polarized mid-infrared radiation (oriented along the antenna´s long axis). **b-d,** Experimental near-field amplitude images, s$_3$ (x, y), of antenna-launched PhPs in $\alpha$-MoO$_3$/SiC at an incident frequency of $\omega_0 = 877$ cm$^{-1}$ (b), $\omega_0 = 870$ cm$^{-1}$ (c) and $\omega_0 = 862$ cm$^{-1}$ (d), respectively. The inset in panel (d) displays a cross section along the black dashed line, showcasing the out-of-plane ray-like propagation of the system. The in-plane propagation occurs highly collimated along the [100] crystal direction, indicative of PhP canalization. **e-g,** Numerically simulated near-field distribution, Re[E$_z$(x, y)], of antenna-launched PhPs in $\alpha$-MoO$_3$/SiC at an incident frequency pf $\omega_0 = 877$ cm$^{-1}$ (e), $\omega_0 = 870$ cm$^{-1}$ (f) and $\omega_0 = 862$ cm$^{-1}$ (g), respectively. Au antennas are demarcated by yellow/black rectangles. The thickness of the $\alpha$-MoO$_3$ flake is 60 nm. **h,** Line profile (black symbols) along the near-field amplitude image in (b) (black dashed line), together with fitted curves following an exponential oscillating function. **i,** Calculated (black solid line) and experimental (red symbols) PhP dispersion

in α-MoO$_3$/SiC. The inset shows the empirical determination of the propagation lifetime of canalized PhPs.

To probe and study experimentally the canalization of PhPs in a single α-MoO$_3$ layer on SiC, we carried out near-field s-SNOM polariton imaging (see Methods) at incident frequencies covering the spectral band where ε$_{SiC}$ and ε$_{optimal}$ take very similar values. For an efficient excitation (launching) of PhPs in the sample, we fabricated a metallic (Au) antenna on its surface (see details in Methods) and illuminated it with s-polarized incident light along its long axis (see schematics in Fig.2a). The near-field images obtained are shown in Fig.2b, 2c, and 2d, corresponding to ω$_0$ = 877cm$^{-1}$, 870 cm$^{-1}$ and 862 cm$^{-1}$, respectively. Excitingly, they all show PhPs emanating from the metallic antenna that propagate directionally along the [100] in-plane crystal direction in α-MoO$_3$. Despite the strong field excited by the antenna and localized at its ends, there are no PhPs launched along the [001] direction. These experimental results thus demonstrate the phenomenon of PhPs canalization emerging in a single α-MoO$_3$ layer, and at several different frequencies, revealing its broadband nature. We note that as a difference to previous visualizations of canalized PhPs[17], the source (antenna) in these experiments is oriented perpendicular to the canalization direction, giving rise to directional plane waves instead of a narrow-collimated beam. To unambiguously verify these findings, we carried out full-wave numerical simulations mimicking the experiments (see Methods). The resulting maps of the out-of-plane electric field component Re(E$_z$) (Fig.2e, 2f and 2g, for ω$_0$ = 877 cm$^{-1}$, ω$_0$ = 870 cm$^{-1}$ and 862 cm$^{-1}$, respectively), display a clear and pronounced directionality of PhPs along the [100] in-plane crystal direction of α-MoO$_3$. This observation qualitatively and quantitatively agrees with the experimental results shown in Fig.2b, Fig.2c, and Fig.2d, respectively. Interestingly, from these simulations we can better understand the canalization phenomenon observed. As shown in the inset in Fig. 2d, PhPs launched by the antenna display out-of-plane ray-like propagation, which is thus restricted to the plane defined by the vertical direction [010] and the in-plane direction [100] of α-MoO$_3$.

Further evidence of the canalization of PhPs can be obtained by analyzing the polaritonic spatial decay. To do this, we extract a line profile from the near-field amplitude images (black symbols in Fig. 2h, corresponding to the dashed line in Fig. 2b), and fit it using an exponentially decaying oscillating function (red curve in Fig. 2h) accounting for both the intrinsic polaritonic damping and its geometrical decay[28]:

$$s(x) = \frac{A}{x^f} e^{-ik_x x} \quad A, k_x, f > 0$$

where $k_x$ is the complex PhP wavevector, A is a fitting parameter, and $f$ is the geometrical decay factor. For the case of hyperbolic propagation in a single α-MoO$_3$ layer[10], $f$ takes values of 0.5, corresponding to the typical geometrical decay of a cylindrical wave. Differently, the resulting curve in Fig.2h yields a value of 0 for $f$, revealing the absence of geometrical decay, and thus underscoring a distinctive feature of PhP canalization[17,19]. More interestingly, the presence of (L=1) polaritonic modes in α-MoO$_3$/SiC leads to a remarkable field confinement factor of λ$_0$/λ$_p$ ~ 48, where λ$_0$ and λ$_p$ represent the free-space light and the polariton wavelengths, respectively. Such factor is notably high for α-MoO$_3$ layers, especially when the thickness is as large as 60 nm, which would typically yield a polaritonic confinement factor of ~7 in a free-standing layer.

Importantly, the broadband nature of canalized PhPs in α-MoO$_3$/SiC allows extracting their group velocity, which has remained elusive so far for any canalized polariton since it typically occurs at a specific frequency (except for canalized PhPs in twisted trilayer structures[19], in which, however, the group velocity has not been studied so far). By extracting the dispersion from near-field measurements at different $\omega_0$ (red dots in Fig.2i), which is in perfect agreement with the dispersion calculated analytically for the L=1 polaritonic mode (black solid line in Fig.2i), we obtain a group velocity $v_g$ of canalized PhPs of ~$10^{-3}c$ at $\omega_0 = 877$ cm$^{-1}$. This small group velocity is similar to that recently measured for PhPs in the hyperbolic regime in $\alpha$-MoO$_3$ (on SiO$_2$ substrate) or in $\alpha$-V$_2$O$_5$[10,11]. In addition, using this $v_g$ and extracting the propagation length $L_p$ from the amplitude line profile (e.g., Fig.2h at $\omega_0 = 877$ cm$^{-1}$), we calculate a lifetime (black spots in the inset of Fig.2i) of ~2 picoseconds, revealing the low loss nature of the canalized PhPs in $\alpha$-MoO$_3$/SiC.

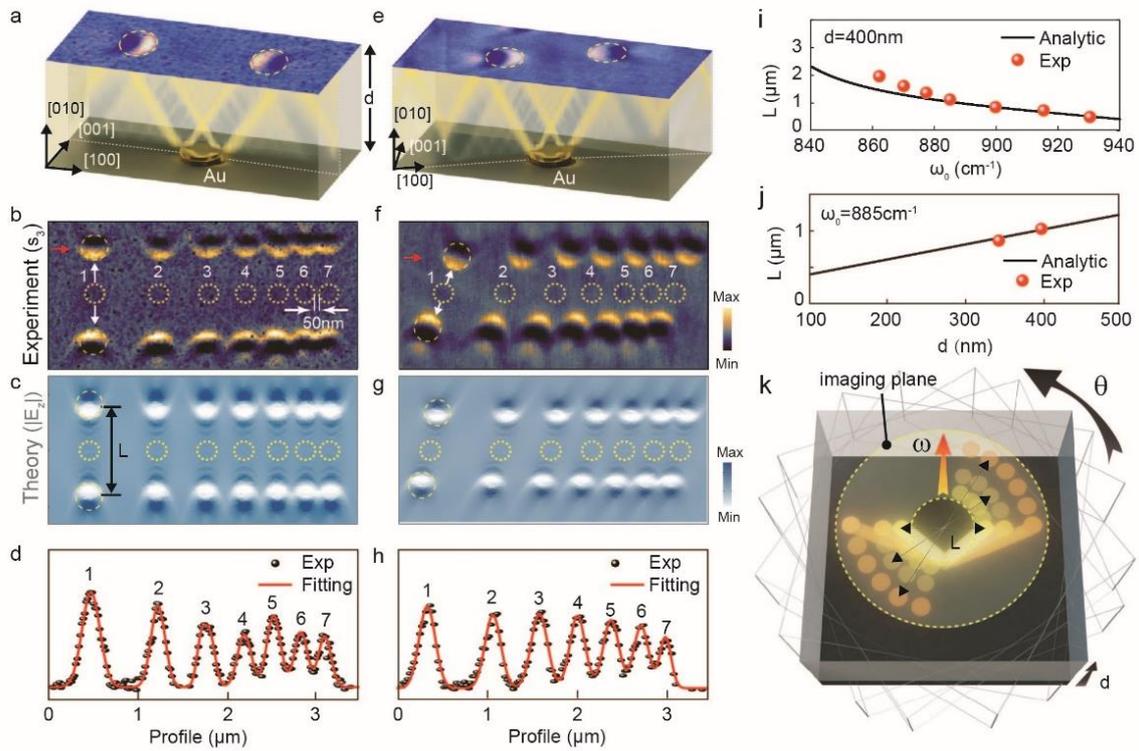

**Figure 3 Super-resolution nano-imaging scheme based on PhP canalization**. **a**, A near-field amplitude image obtained at the uppermost surface of $\alpha$-MoO$_3$/Au disk/SiC. Calculations of the intensity of the out-of-plane component of the electric field, $|E_z|$ are also shown in false color. The 200 nm diameter gold disks beneath the $\alpha$-MoO$_3$ layer are encircled in yellow. All data were obtained at $\omega_0 = 885$ cm$^{-1}$, with an $\alpha$-MoO$_3$ layer thickness d =415 nm. **b-c**, Near-field amplitude image (b) and simulated near-field intensity (c) obtained at the top surface of $\alpha$-MoO$_3$ layers on a row of seven gold disks (labelled as 1-7), whose edge-to-edge horizontal separations are 500, 300, 200, 150, 100, and 50 nm, respectively. **d**, Near-field intensity profiles (black symbols) extracted along the horizontal direction marked by the red arrows in (b). The red solid lines represent fitting curves by a Lorentz function. **e-h**, the same images as (a-d) but the $\alpha$-MoO$_3$ layer is rotated 25° clockwise relative to the one in (a-d). **i-j**, Dependence of the separation between two images of the gold disks (denoted by L in the panel (c)) with the incident frequencies (i) and thicknesses of $\alpha$-MoO$_3$ flake (j). The continuous lines in (i) and (j) are extracted from analytical calculations and the symbols, from experimental s-SNOM images. **k**, Schematic of super-resolution imaging scheme capable of delivering the super-

resolved projection of Au disks to the remote locations on the top plane. By varying the incident frequency, the thicknesses of α-MoO$_3$ flakes, and rotating the α-MoO$_3$ flake, we can obtain images at desired locations.

The low losses together with the broadband response of canalized PhPs in individual α-MoO$_3$ layers open the door for applications involving deeply subwavelength-scale information transfer. For illustrative purposes, we demonstrate in the following super-resolution imaging of buried gold nanodisks using a single layer of α-MoO$_3$ as a superlens capable of leveraging diffraction-less polariton propagation to image nanometer-sized objects. To that end, we fabricate a gold disk with a diameter of 200 nm on the surface of a SiC substrate and cover it with an α-MoO$_3$ layer. By measuring the top α-MoO$_3$ surface with s-SNOM (Fig. 3a), we observe a pair of bright-dark semicircular patterns, which reflect the typical near-field distribution of a dipolar mode in a metallic antenna[29]. More importantly, the overall size of these near-field patterns is similar to that of the gold disk, allowing thus to obtain a reliable projection of the buried element on the top surface. We explain this result by the unique combination of both out-of-plane hyperbolic (ray-like) and in-plane diffraction-less propagation (canalization) of PhPs in α-MoO$_3$ when it is placed on SiC (see cross-section in Fig.3a), which ensures the electromagnetic field energy to remain highly confined within the x-z plane defined by the [010] and [100] crystal directions in α-MoO$_3$. Note that this imaging scenario is very different from that reported in previous nano-imaging studies using strongly confined polaritons, in which the combination of out-of-plane hyperbolic and in-plane isotropic propagation (as e.g., PhPs in h-BN)[30,31] yield images of buried disks on the top plane exhibiting more complex patterns formed by enlarged circles along all in-plane directions. Consequently, we can consider a single layer of α-MoO$_3$ on SiC as a superlens that, via PhPs canalization, allows direct projection of the contour of buried elements at two locations laterally displaced (from the source vertical) in the top plane.

Having established such a principle of nano-imaging, we delve now into its capability in terms of lateral resolution. To that end, we fabricate a row of seven Au disks (yellow dashed circles in Fig.3b-3c) positioned with incrementally increasing edge-to-edge separations of 50, 100, 150, 200, and 500 nm. The near-field amplitude image (Fig.3b), taken at an incident wavelength of $\lambda_0$ = 11.3 μm (885 cm$^{-1}$), shows that even the adjacent disks with a mere 50 nm spacing (labelled as 6 and 7 in Fig.3b) are resolved on the top plane. The corresponding simulated image mimicking the experiment corroborates this result (Fig. 3c). To quantitatively analyze the resolution achieved in these experiments, we extract a line profile (Fig.3d) from the near-field amplitude image along the alignment direction of the disks (red arrow in Fig.3b). We clearly observe seven peaks (Fig.3d) corresponding to the seven buried Au disks, which thus indicates the capability to distinguish separations of about 50 nm, corresponding to a subdiffractional resolution of ~$\lambda_0$/220. A comparison with an optical scheme in which the imaging is obtained by employing PhPs with a typical hyperbolic propagation, i.e., without canalization, is shown in Fig. S12, where only 4 peaks are observed, indicating a much lower lateral resolution.

An interesting property of our canalization-based nanoimaging is that the locations on the surface where the buried objects are projected are tunable by a mere rotation of the individual α-MoO$_3$ layer with respect to the SiC substrate. As an example, the near-field images in Figs. 3e and 3f show the case for a rotation of the α-MoO$_3$ flake of 25° (using the [010] crystal direction as the rotation axis). It can be clearly seen that the Au disks are now located on the surface at different positions, resulting from an in-plane polar translation of 25°, in excellent agreement with the corresponding simulated image (Fig. 3g). Importantly, by extracting a line profile (Fig.3h) from the near-field amplitude image (indicated by the red arrow in Fig.3f), we observe that despite the translation, the spatial resolution achieved is the same as in the original case (Fig. 3d). This result suggests that given the possibility to rotate the α-MoO$_3$ layer to any angle in the plane, the projected images in our canalization-based nanoimaging scheme can be adjusted to any in-plane location following a circumference (indicated by dashed curves in the schematic of Fig. 3k). Furthermore, since the diameter of the circumference (denoted by L in Fig. 3c and 3k) can in turn be tuned by parameters that modify the angle of propagation of the out-of-plane hyperbolic rays, such as the frequency (Fig. 3i) or the thickness of the α-MoO$_3$ layer (Fig. 3j), the position of the projected image on the surface can be potentially adjustable over wide regions. To evaluate this possibility, we calculated L as a function of $\omega_0$ (solid black line in Fig. 3i) and the layer thickness d (solid black line in Fig. 3j), obtaining as a result a broad tunability (from ~0.5 μm to ~2 μm) with both parameters. To experimentally corroborate these theoretical results, we, therefore fabricated two α-MoO$_3$/SiC superlenses in which the α-MoO$_3$ layer thickness was different ($d_1$ = 415 nm and $d_2$ = 340nm) and measured by s-SNOM the images projected by buried Au disks at different incident frequencies. The L values (red symbols in Fig.3i-3j) extracted from these images (Fig. S10-S11), show clear quantitative agreement with the theoretical curves, thus underlining the pioneering capabilities of the imaging scheme based on PhPs canalization in individual α-MoO$_3$ layers to produce super-resolved images at desired locations on the surface (Fig.3k).

Overall, we demonstrate a novel type of PhP canalization present in individual α-MoO$_3$ thin layers that not only streamlines the sample fabrication compared to twisted stacks but also exhibits a broad spectral coverage (from $\omega_0$ = 850 cm$^{-1}$ to $\omega_0$ = 900 cm$^{-1}$) and low optical losses (lifetimes of ~2 picoseconds). As a result of these unique properties, we showcase a first proof-of-concept application of PhPs canalization: super-resolved imaging of buried nanometer-sized objects (Au disks) at desired locations on the upper surface, enabling resolutions as high as ~$\lambda_0$/220. Altogether, our results open new avenues for integrated flat optics, nanoscale information transfer, and heat management applications employing highly collimated PhPs.